\documentclass[a4paper,11pt]{article}
\pdfoutput=1 

\usepackage{jcappub} 

\usepackage[T1]{fontenc} 

\usepackage[margin=0.9in]{geometry}

\usepackage{amsmath,amssymb}
\usepackage[utf8]{inputenc}
\usepackage{graphicx}
\usepackage{subcaption}
\usepackage[font=footnotesize]{caption}
\usepackage[colorlinks=true, linkcolor=blue, citecolor=blue, urlcolor=blue]{hyperref}
\usepackage{tabularx}

\title{The Noise of Vacuum}

\author{Gabriela Barenboim}
\emailAdd{gabriela.barenboim@uv.es}
\affiliation{Departament de F\'isica Te\`orica and IFIC, \\
Universitat de Val\`encia-CSIC, E-46100, Burjassot, Spain}

\date{\today}

\abstract{
We investigate the evolution of primordial cosmological perturbations in a vacuum decay model where de Sitter space transitions to radiation domination through quantum-thermal decay processes. Unlike standard inflation, this framework generates curvature perturbations through stochastic noise from vacuum decay rather than quantum fluctuations of an inflaton field. We derive the stochastic differential equation governing the curvature perturbation $\mathcal{R}(t)$ and show that any horizon crossing is brief and does not constitute the primary mechanism for perturbation generation. Scale dependence emerges from spatial correlations in the noise rather than horizon crossing dynamics. The model naturally addresses the horizon and flatness problems through initial thermal equilibrium in de Sitter space and predicts zero tensor-to-scalar ratio. We demonstrate that spatially correlated noise can generate observationally viable spectral tilts while maintaining Gaussian statistics.
}

\begin{document}

\maketitle

\flushbottom 

\section{Introduction}
The inflationary paradigm has been remarkably successful in explaining the observed homogeneity, isotropy, and flatness of our universe, as well as providing a mechanism for generating primordial density perturbations \cite{Guth:1981, Linde:1982, Albrecht:1982}. However, the framework relies on the introduction of one or more scalar fields (inflatons) with carefully tuned potentials, raising questions about naturalness and predictability \cite{Martin:2013}.

While vacuum decay scenarios in cosmology have been explored previously \cite{Berera:1995wh,Ford:1987}, our framework differs fundamentally through a frame-dependent thermal interpretation of de Sitter space based on global satisfaction of the Kubo-Martin-Schwinger (KMS) condition in the cosmic rest frame. Recent work by Alicki et al. \cite{Alicki:2023rfv} has argued for a refined understanding of de Sitter thermality in which the KMS condition---the rigorous quantum statistical criterion for thermal equilibrium---is globally satisfied only in the \emph{cosmic rest frame} defined by observers comoving with the FLRW expansion. 
In this interpretation, quantum fields in de Sitter space behave as a genuine thermal bath (in the sense of satisfying the KMS condition globally in the cosmic rest frame, as detailed in Ref.~\cite{Alicki:2023rfv}) at the Gibbons-Hawking temperature $T_{\rm dS} = h/(2\pi)$, with particle creation and annihilation processes maintaining statistical balance.

Clearly,  this represents a specific theoretical framework that goes beyond the standard horizon-dependent notion of de Sitter temperature, which does not imply global thermal equilibrium, a time-independent density matrix, or a well-defined Hamiltonian generating time translations.
While this interpretation remains subject to ongoing theoretical investigation, it provides a self-consistent foundation for treating vacuum decay as a thermodynamic process, which we explore phenomenologically in this work. The thermal interpretation of de Sitter space, originally established by Gibbons and Hawking \cite{Gibbons:1977}, requires careful consideration of observer frames, and the framework proposed in \cite{Alicki:2023rfv} offers a specific resolution to these frame-dependent subtleties.

Unlike static patch observers who experience only local thermal properties \cite{Bousso:2002}, comoving observers in the cosmic rest frame see the universe as homogeneous and isotropic, experiencing a thermodynamically consistent environment characterized by the de Sitter temperature. This distinction is crucial because it allows us to treat the early universe as beginning in a well-defined thermal state, with vacuum decay processes governed by standard quantum statistical mechanics.

Building on this thermodynamic foundation, we explore a model where the early universe begins in this thermal de Sitter state and transitions to radiation domination through quantum decay processes. The statistical mechanics of this decay, combined with the thermal properties specific to the cosmic rest frame, provides a natural framework for generating primordial perturbations without requiring additional scalar fields.

This decay generates stochastic fluctuations that seed primordial perturbations, offering a potentially simpler explanation for observed cosmological features. The model leverages established physics---specifically, the frame-dependent thermodynamic properties of de Sitter space and quantum statistical mechanics---without introducing exotic fields or potentials.

Our approach differs fundamentally from standard inflation in several key aspects: (i) perturbations arise from stochastic vacuum decay rather than inflaton fluctuations, (ii) horizon crossing is typically brief and not the dominant perturbation mechanism, (iii) scale dependence emerges from noise correlations rather than horizon crossing dynamics, and (iv) the model predicts zero gravitational waves. The thermal interpretation, properly understood in the cosmic rest frame, provides the physical foundation for this alternative paradigm.

\section{The Vacuum Decay Model}
Following the framework introduced in \cite{Alicki:2023rfv,Alicki:2023tfz}, we consider a Universe in which de Sitter vacuum energy behaves as a thermal reservoir when viewed from the cosmic rest frame---the natural reference frame for cosmological observers comoving with the FLRW expansion. In this frame, the quantum properties of de Sitter space satisfy the Kubo-Martin-Schwinger condition globally, establishing what we refer to as true thermal equilibrium (in the specific sense defined in Ref.~\cite{Alicki:2023rfv}) at the Gibbons-Hawking temperature $T_{\rm dS} = h/(2\pi)$ rather than the observer-dependent thermal effects experienced in other reference frames. The decay of vacuum energy into radiation is treated as an irreversible process governed by open quantum system dynamics within this thermodynamically consistent framework. This allows for a smooth and natural transition from an inflationary phase to a radiation-dominated era, without the need for an explicit reheating mechanism or an inflaton field.

Our analysis relies on the frame-dependent 
thermal interpretation of de Sitter space developed in \cite{Alicki:2023rfv}, 
wherein the cosmic rest frame admits a globally defined thermal state satisfying 
the KMS condition. This framework differs from standard treatments in several 
respects: (i) it posits genuine thermal equilibrium for comoving observers rather 
than merely observer-dependent thermal effects, (ii) it provides a thermodynamic 
foundation for vacuum decay absent in conventional de Sitter quantum field theory, 
and (iii) it enables statistical mechanical treatment of early universe dynamics. 
We acknowledge this interpretation is not universally accepted and represents a 
theoretical choice with specific assumptions about quantum statistical mechanics 
in curved spacetime. The viability of the resulting phenomenological framework 
will ultimately be judged by its observational predictions ($r \approx 0$, 
Gaussian perturbations) rather than by settling foundational debates about 
de Sitter thermality.

This frame-dependent thermal interpretation is illustrated schematically in Fig.~\ref{fig:temperature_perception} (see Appendix A for a detailed discussion). Static patch observers, confined to causal patches within de Sitter space, experience position-dependent local temperatures $T_{\rm loc} = h/(2\pi\sqrt{1-h^2r^2})$ due to gravitational redshift effects, where the temperature varies with position relative to the horizon. In contrast, comoving observers in the cosmic rest frame experience a uniform global temperature $T = h/(2\pi)$ throughout space, reflecting the global satisfaction of the KMS condition in this frame. This distinction is fundamental to our model, as it establishes the thermodynamic consistency required for treating vacuum decay as a well-defined statistical process

We now describe the background evolution of this vacuum decay model.\footnote{Throughout this article we work in Planck units, such that $G = M_{\rm Pl}^{-2} = \hbar = k_B = c = 1$.}
 
\begin{figure}[h]
\centering
\includegraphics[width=0.7\textwidth]{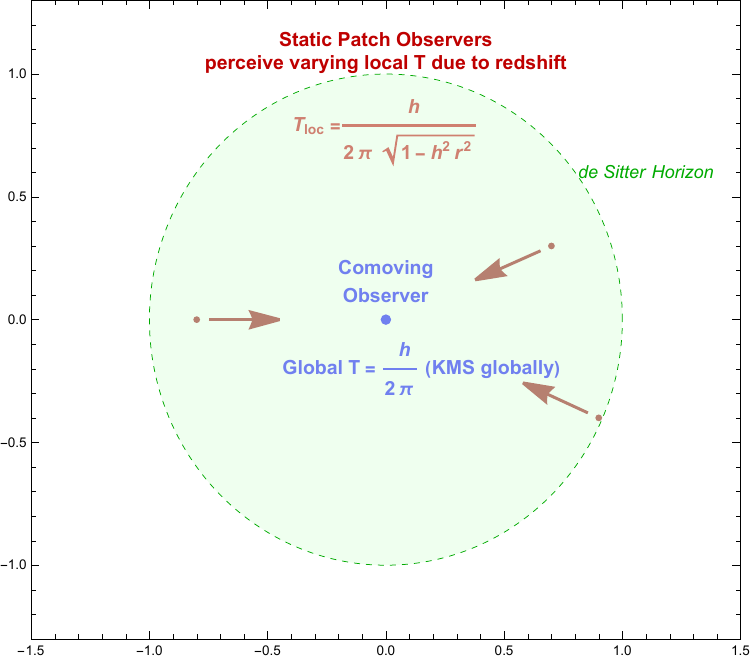}
\caption{Schematic illustration of temperature perception in de Sitter space. Static patch observers experience position-dependent local temperature due to redshift effects, while comoving observers in the cosmic rest frame experience a global temperature $T = h/(2\pi)$ where the KMS condition is satisfied globally.}
\label{fig:temperature_perception}
\end{figure}

\subsection{Background Evolution}
We consider a universe containing both de Sitter vacuum energy and radiation, with energy densities:
\begin{equation}
\rho_{\rm dS}(t) = \sigma h(t)^4 
\label{eq:rho_dS}
\end{equation}\footnote{Other functional dependences are conceivable; the present choice represents the minimal form consistent with dimensional analysis and the assumed thermal scaling $T_{\rm dS} \sim h$, leading to $\rho \sim T^4 \sim h^4$.}
\begin{equation}
\rho_r(t) = \frac{3}{8\pi} h(t)^2 - \sigma h(t)^4 
\label{eq:rho_r}
\end{equation}
These equations represent phenomenological parametrizations of the energy content during the vacuum decay transition. Equation~\eqref{eq:rho_dS} models the vacuum energy density with a $h^4$ scaling, which can be motivated thermodynamically: in the thermal interpretation of \cite{Alicki:2023rfv}, the Gibbons-Hawking temperature scales as $T_{\rm dS} \sim h$, and a thermal energy density would scale as $\rho \sim T^4 \sim h^4$. The coefficient $\sigma$ parameterizes the effective number of degrees of freedom contributing to this thermal vacuum energy. Equation~\eqref{eq:rho_r} then ensures that the total energy density satisfies the Friedmann equation $\rho_{\rm eff} = \frac{3}{8\pi}h^2$, with the radiation component defined as the residual after subtracting the vacuum contribution.

We acknowledge these are effective parametrizations rather than microscopic derivations; 
the $h^4$ scaling for vacuum energy follows naturally from thermal considerations
($\rho \sim T^4$ with $T \sim h$), while Eq.~\eqref{eq:rho_r} ensures consistency 
with the Friedmann equation. These choices provide the simplest phenomenological 
realization of vacuum decay within the thermal framework, analogous to how inflaton 
potentials parametrize inflation without full microscopic justification.

Here $h(t)$ denotes the Hubble parameter in Planck units, which with our convention ($M_{\rm Pl} = 1$) is simply $h(t) = H(t) = \dot{a}/a$ with dimensions of $[M_{\rm Pl}]^{-1} = 1$. The parameter $\sigma$ is a constant related to the number of massless degrees of freedom. The parameter $\sigma$ affects only the timescales and critical values but not the essential phenomenology. For concreteness, we set $\sigma = 3/(8\pi)$ throughout this work, which normalizes the critical equilibrium value $h_0 \equiv \sqrt{3/(8\pi\sigma)} = 1$. The energy density of the regular particle content of the Universe is expressed as
\[
\rho(t) = \frac{3}{8\pi} h(t)^2 - \sigma h(t)^4,
\]
combining the contributions from both radiation and the de Sitter thermal bath.

The evolution equation for the Hubble parameter, derived from energy-momentum conservation with vacuum decay, is:
\begin{equation}
\dot{h}(t) = -\frac{16\pi\sigma}{3} h(t)^2 \left(\frac{3}{8\pi\sigma} - h(t)^2\right) 
\label{eq:hubble_evolution}
\end{equation}

This equation describes the transition from de Sitter domination to radiation domination, with the critical value $h_c = \sqrt{3/(16\pi\sigma)}$ marking the transition point. Importantly, the initial vacuum state before decay does not require additional assumptions; any small fluctuation suffices to trigger the vacuum decay process, making the model robust to the choice of initial conditions.

Strictly speaking,  our model describes a \emph{transition} from an initial state with $h(t) \approx h_0$ (nearly constant) to radiation domination with decreasing $h(t)$. Strictly speaking, once $h$ becomes time-dependent, the spacetime is no longer pure de Sitter but quasi-de Sitter. We use ``de Sitter vacuum energy'' and ``thermal de Sitter state'' when referring to the initial conditions and the thermodynamic framework of \cite{Alicki:2023rfv}, which was developed for constant-$h$ de Sitter space. The applicability of this thermal interpretation during the transition phase (where $h$ varies) is an assumption of our phenomenological framework, as discussed in Sec.~\ref{sec:limitations}.

\begin{figure}[h!]
    \centering
    \includegraphics[width=0.7\textwidth]{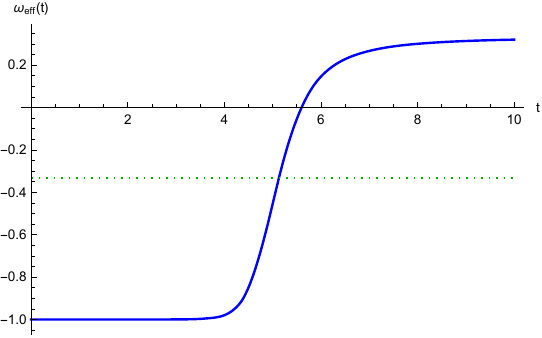}
  \caption{Effective equation of state $\omega_\mathrm{eff}(t)$ as a function of cosmic time $t$. The dotted green line indicates the value of $\omega_\mathrm{eff}$ corresponding to equality between vacuum and radiation. Time is measured in units of the Planck time.}
    \label{fig:weff}
\end{figure}

To characterize the combined fluid, the effective equation of state $w_{\rm eff}(t)$ is introduced:
\[
w_{\rm eff}(t) = \frac{p_{\rm dS} + p_r}{\rho_{\rm dS} + \rho_r},
\]
where $p_{\rm dS} = -\sigma h(t)^4$ and $p_r = \frac{1}{3} \rho_r(t)$ are the pressures for de Sitter space and radiation, respectively. Substituting for $\rho_r$ and $p_{\rm dS}$, the effective equation of state simplifies to:
\begin{equation}
w_{\rm eff}(t) = -\frac{32\pi}{9} \sigma h(t)^2 + \frac{1}{3}.
\end{equation}
This expression captures the time-dependent nature of the equation of state as the Universe evolves, reflecting the changing contributions from the de Sitter thermal bath and the radiation field. The analysis shows that the Universe naturally transitions from inflation to a radiation-dominated phase, with the effective equation of state evolving from a nearly constant $-1$ during inflation to a value closer to $\frac{1}{3}$ as radiation becomes dominant, see Figure~\ref{fig:weff}.

\subsection{Mathematical Formulation of Thermal Equilibrium}
\label{sec:thermal_math}

Following Ref.~\cite{Alicki:2023rfv}, we now make precise the mathematical framework underlying the thermal interpretation of de Sitter space. This involves three key elements: (i) a Markovian master equation describing how localized quantum systems evolve in the expanding background, (ii) the Kubo-Martin-Schwinger (KMS) condition that characterizes thermal equilibrium, and (iii) the frame-dependent nature of this thermality. Readers primarily interested in the cosmological phenomenology may skip to Sec.~3, noting that the key result is the emergence of a thermal bath at temperature $T_{\rm dS} = h/(2\pi)$ with energy density $\rho_{\rm dS} \propto h^4$.

\subsubsection{System-Environment Decomposition and the Markovian Master Equation}

Consider a localized quantum system (characterized by bare Hamiltonian $\hat{H}_0$) embedded in de Sitter space and weakly coupled to a massless quantum field. The dynamics in expanding space are governed by the modified Hamiltonian:
\begin{equation}
\hat{H}_D(t) = \hat{H}_0 + h(t)\hat{D},
\label{eq:H_dilation}
\end{equation}
where $\hat{D}$ is the spatial dilation operator and $h(t)$ is the Hubble parameter. For a wave function $\psi(\mathbf{x})$, the dilation operator is defined through:
\begin{equation}
\left[e^{-i\lambda\hat{D}}\psi\right](\mathbf{x}) = e^{-\frac{3}{2}\lambda}\psi\left(e^{-\lambda}\mathbf{x}\right),
\end{equation}
which yields 
\begin{equation}
\hat{D} = \frac{1}{2}(\hat{\mathbf{x}}\cdot\hat{\mathbf{p}} + \hat{\mathbf{p}}\cdot\hat{\mathbf{x}}).
\end{equation}

The coupling to the massless field background is described by an interaction Hamiltonian:
\begin{equation}
\hat{H}_{\rm int} = \lambda\left(\hat{b} + \hat{b}^{\dagger}\right) \otimes \hat{\phi}_\Lambda(0),
\end{equation}
where $\hat{b},\hat{b}^{\dagger}$ are ladder operators for the localized quantum system (e.g., harmonic oscillator modes), $\lambda$ is a dimensionless coupling constant characterizing the interaction strength, and $\hat{\phi}_\Lambda(0)$ represents the regularized quantum field at the system's location with ultraviolet cutoff parameter $\Lambda$. In the weak-coupling regime ($\lambda \ll 1$), the Markovian approximation is valid.

Under standard assumptions of weak coupling and separation of timescales between the system and the bath's internal relaxation, one derives the Markovian master equation (MME) for the reduced density matrix $\hat{\rho}(t)$ of the system \cite{Alicki:2023rfv}. Here $\hat{H}$ denotes the renormalized system Hamiltonian after accounting for expansion effects (see Appendix of Ref.~\cite{Alicki:2023rfv} for the relationship between bare and renormalized Hamiltonians):
\begin{equation}
\frac{d}{dt}\hat{\rho}(t) = -i\left[\hat{H},\hat{\rho}(t)\right] + \frac{1}{2}\left(\gamma_\downarrow\left(\left[\hat{b},\hat{\rho}(t)\hat{b}^{\dagger}\right] + \left[\hat{b}\hat{\rho}(t),\hat{b}^{\dagger}\right]\right) + \gamma_\uparrow\left(\left[\hat{b}^{\dagger},\hat{\rho}(t)\hat{b}\right] + \left[\hat{b}^{\dagger}\hat{\rho}(t),\hat{b}\right]\right)\right),
\label{eq:MME}
\end{equation}
where $\gamma_\downarrow = \lambda^2\tilde{G}(\omega)$ and $\gamma_\uparrow = \lambda^2\tilde{G}(-\omega)$ are damping and pumping rates determined by the spectral density $\tilde{G}(\omega)$—the Fourier transform of the bath correlation function $G(t) = \langle\hat{\phi}(t)\hat{\phi}(0)\rangle$.

\subsubsection{The KMS Condition and Thermal State}

The key result of Ref.~\cite{Alicki:2023rfv} is that for constant Hubble parameter ($h = \text{const.}$), the spectral density satisfies the Kubo-Martin-Schwinger (KMS) condition:
\begin{equation}
\tilde{G}(-\omega) = e^{-\omega/T_{\rm dS}}\tilde{G}(\omega),
\label{eq:KMS}
\end{equation}
where $T_{\rm dS} = h/(2\pi)$ is the Gibbons-Hawking temperature. Equivalently, writing the inverse temperature as $(T_{\rm dS})^{-1} = 2\pi/h$, this becomes $\tilde{G}(-\omega) = \exp(-2\pi\omega/h)\,\tilde{G}(\omega)$. This is the rigorous mathematical criterion for thermal equilibrium in quantum statistical mechanics.

This KMS condition implies that the Gibbs state:
\begin{equation}
\hat{\rho}_{\rm eq} = \frac{1}{Z}e^{-\hat{H}/T_{\rm dS}}, \quad Z = \text{Tr}\left(e^{-\hat{H}/T_{\rm dS}}\right),
\label{eq:Gibbs}
\end{equation}
is an equilibrium state for the localized system, and moreover that any initial state relaxes to $\hat{\rho}_{\rm eq}$.

The energy density of this thermal bath obeys the Stefan-Boltzmann law:
\begin{equation}
\rho_{\rm dS} = \int_0^\infty d\omega\, \mathcal{U}(\omega) = \int_0^\infty d\omega\, \frac{n(\omega)\omega}{e^{\omega/T_{\rm dS}}-1} \propto T_{\rm dS}^4 = \frac{h^4}{(2\pi)^4},
\label{eq:StefanBoltzmann}
\end{equation}
where $n(\omega) = \omega^2/(2\pi^2)$ is the density of states for massless bosons and $\mathcal{U}(\omega)$ is the energy density spectrum (the Planck distribution per polarization). The proportionality constant depends on the effective number of field degrees of freedom contributing to the thermal bath; in our phenomenological model (Sec.~2.1), we parametrize this as $\rho_{\rm dS} = \sigma h^4$ where $\sigma$ is a constant encoding this effective degree-of-freedom count.

\subsubsection{From Quantum to Classical Stochastic Description}

The MME in Eq.~\eqref{eq:MME} describes the quantum evolution of a localized system's density matrix. In our cosmological application, the curvature perturbation $\mathcal{R}$ represents a semiclassical field whose quantum fluctuations have decohered to classical stochastic behavior through environmental interactions \cite{Starobinsky:1986fx, Calzetta:2008}. The transition from the quantum MME to the classical Langevin equation in Eq.~\eqref{eq:langevin} follows standard procedures: one takes expectation values of the relevant observables, and the quantum noise operators become classical stochastic sources with correlation functions determined by the KMS condition. This quantum-to-classical transition is well-established in the stochastic inflation literature \cite{Starobinsky:1986fx} and in open quantum systems approaches to cosmology \cite{Calzetta:2008}. The essential physics—thermal noise with KMS correlations sourcing perturbation growth—is preserved in this classical limit.

In our cosmological application, the relevant coupling is between curvature perturbations and the decaying vacuum energy; we do not compute the microscopic coupling $\lambda$ from first principles but incorporate its effects phenomenologically through the noise amplitude $\beta(t)$ in Eq.~\eqref{eq:langevin}.

\subsubsection{Frame Dependence and Scope of Applicability}

\textbf{Critical Assumption:} The mathematical framework presented above—including the KMS condition (Eq.~\ref{eq:KMS}), thermal state (Eq.~\ref{eq:Gibbs}), and Stefan-Boltzmann law (Eq.~\ref{eq:StefanBoltzmann})—was rigorously derived in Ref.~\cite{Alicki:2023rfv} for pure de Sitter space with strictly constant Hubble parameter $h = \text{const.}$ Our vacuum decay model involves time-dependent $h(t)$, meaning the spacetime is quasi-de Sitter rather than pure de Sitter. The applicability of the thermal interpretation during this time-dependent phase is an \emph{assumption} that we adopt as a working hypothesis for phenomenological exploration, not a derived result. This represents an extrapolation beyond the regime where the framework is rigorously established. We return to this limitation in Sec.~\ref{sec:limitations}.

Additionally, this thermal interpretation is frame-dependent \cite{Alicki:2023rfv}. The KMS condition and the associated temperature $T_{\rm dS}$ are well-defined only in the cosmic rest frame—the frame in which the FLRW metric takes the form $ds^2 = -dt^2 + a^2(t)d\mathbf{x}^2$ with comoving spatial coordinates $\mathbf{x}$. Observers moving with respect to this frame would not measure the same thermal properties. This is analogous to how the cosmic microwave background has a well-defined blackbody temperature only in the frame where it has no dipole moment.

\subsection{Effective Sound Speed}
The effective sound speed $c_s^2(t)$ characterizes the response of pressure perturbations to density perturbations in the mixed fluid. For a perfect fluid, it is defined as \cite{Weinberg:2008zzc}:
\begin{equation}
c_s^2(t) = \frac{\delta p}{\delta \rho} = \frac{p_{\rm eff}'(t)}{\rho_{\rm eff}'(t)}
\end{equation}
where the prime denotes time derivative. The total energy density and pressure are:
\begin{align}
\rho_{\rm eff}(t) &= \rho_{\rm dS}(t) + \rho_r(t) = \frac{3}{8\pi} h(t)^2 \\
p_{\rm eff}(t) &= p_{\rm dS}(t) + p_r(t) = -\sigma h(t)^4 + \frac{1}{3}\left(\frac{3}{8\pi} h(t)^2 - \sigma h(t)^4\right)
\end{align}

Computing the time derivatives and simplifying yields:
\begin{equation}
c_s^2(t) = \frac{4\pi}{3}\left(-\frac{16}{3} \sigma h(t)^2 + \frac{1}{4\pi}\right) 
\end{equation}

The negative value during early times reflects the instability of the de Sitter vacuum, signaling the natural decay process that drives the universe toward radiation domination. This negative sound speed is not unphysical but rather indicates that the system is unstable and evolving away from the initial de Sitter state.

\subsection{Distinction from Bubble Nucleation and Old Inflation}

Our vacuum decay model shares superficial similarities with Guth's original ``old inflation'' paradigm \cite{Guth:1981}, particularly in featuring a de Sitter phase transitioning to radiation domination. However, the physical mechanisms differ fundamentally, and our approach avoids the pathologies that plagued old inflation.

In old inflation, the false vacuum decays via \emph{bubble nucleation}---a first-order phase transition where causally disconnected regions undergo tunneling at random spacetime points. The exponential expansion ensures these bubbles never percolate, resulting in a universe dominated by false vacuum regions separated by rare bubble walls. This is the infamous ``graceful exit problem'' \cite{Guth:1981,Linde:1982}: the transition never completes globally, and perturbations from discrete bubble placement and colliding walls produce unacceptably large inhomogeneities \cite{Hawking:1982}.

In contrast, our model describes a \emph{continuous, homogeneous} decay process. The key distinction lies in the frame-dependent thermal interpretation of de Sitter space \cite{Alicki:2023tfz}. For comoving observers in the cosmic rest frame, quantum fields satisfy the Kubo-Martin-Schwinger (KMS) condition globally, establishing true thermal equilibrium at the Gibbons-Hawking temperature $T_{\rm dS} = h/(2\pi)$ throughout all space. Every comoving patch experiences statistically identical conditions, with vacuum decay occurring as a \emph{uniform statistical process} everywhere simultaneously, driven by the same thermal ensemble.

The microphysical mechanisms differ correspondingly. Old inflation proceeds via quantum tunneling with rate $\Gamma/V \propto e^{-S_{\rm CDL}}$ determined by the Coleman-De Luccia action \cite{Coleman:1980}, where bubbles nucleate randomly and their interiors expand. Our thermal vacuum decay occurs through continuous energy-momentum transfer from the thermal de Sitter vacuum to radiation, mediated  by gravitational interactions and governed by open quantum system dynamics \cite{Alicki:2023tfz}, wherein microscopic quantum degrees of freedom are traced out to yield an effective stochastic description for macroscopic cosmological perturbations (detailed in Sec.~3.2).

The decay rate is determined by thermal equilibrium statistical mechanics rather than tunneling amplitudes, and occurs uniformly because the thermal state is global in the cosmic rest frame.

Our model avoids the graceful exit problem entirely because there are no bubbles. The Hubble parameter $h(t)$ evolves according to Eq.~\eqref{eq:hubble_evolution}, describing a smooth, deterministic transition from de Sitter domination to radiation. The homogeneity of the final state is guaranteed by the symmetries of the initial thermal de Sitter state, not by bubble percolation or fine-tuning decay rates.

The horizon problem is also resolved differently. In old inflation, bubble nucleation produces uncorrelated domains, exacerbating the problem. Our framework resolves this through \emph{statistical equivalence}: all regions begin in the same global thermal equilibrium, with homogeneity reflecting the initial state's symmetries rather than causal processes during inflation. Perturbations arise from quantum fluctuations in the rate of energy-momentum transfer during uniform decay, with amplitudes naturally of order $\mathcal{O}(10^{-5})$ for appropriate parameters, compatible with CMB observations.

\medskip

\noindent\textbf{Note on Fields:} While our model does not require a new fundamental scalar field with a carefully tuned potential (the inflaton of standard models), the vacuum decay process does involve quantum fields---specifically, the gravitational field and Standard Model degrees of freedom. The key distinction is that we do not postulate exotic fields or potentials beyond known physics. The effective description through open quantum systems captures the relevant dynamics without requiring new microphysical assumptions about inflaton couplings or reheating mechanisms. 

However, we acknowledge that assumptions about Standard Model field behavior at early 
times remain implicit: (i) thermalization of produced radiation must occur on timescales 
$\lesssim H^{-1}$, (ii) coupling mechanisms between vacuum energy and SM degrees of 
freedom (e.g., gravitational particle creation, conformal anomaly effects) must be 
sufficiently strong to drive the decay rate in Eq.~\eqref{eq:hubble_evolution}, and 
(iii) the effective parametrizations in Eqs.~\eqref{eq:rho_dS}--\eqref{eq:rho_r} 
encode coarse-grained microphysical interactions. The key distinction from inflation 
is that these assumptions involve known field content and established gravitational 
physics rather than exotic scalar fields with fine-tuned potentials.

\section{Stochastic Evolution of Curvature Perturbations}
\subsection{The Stochastic Framework}
The curvature perturbation $\mathcal{R}(t)$, which tracks the evolution of comoving slices of constant energy density, becomes sensitive to the microphysical details of the vacuum decay. This decay introduces not only deterministic evolution but also an element of randomness due to quantum fluctuations associated with the decay itself.

As a result, the evolution of $\mathcal{R}(t)$ is governed by a stochastic differential equation (SDE) of the form\footnote{This first-order approximation is valid for sub-Hubble modes where the oscillatory term $c_s^2 \nabla^2 \mathcal{R}$ dominates over the source term. The accuracy of this simplification is confirmed through numerical integration of the complete second-order evolution equation (see Appendix B for details).}
\begin{equation}
\dot{\mathcal{R}}(t) + \alpha(t) \mathcal{R}(t) = \beta(t) \xi(t),
\label{eq:langevin}
\end{equation}

where:
\begin{itemize}
    \item $\alpha(t)$ is a time-dependent damping coefficient that quantifies the rate at which the curvature perturbation is dissipated due to expansion and interactions within the mixed fluid. It is given by:
\[
\alpha(t) = \left(1 + w_{\rm eff}(t)\right) h(t),
\]
where $w_{\rm eff}(t)$ is the effective equation of state parameter obtained before. Explicitly,
\begin{equation}
\alpha(t) = h(t) \left(\frac{4}{3} - \frac{32\pi}{9} \sigma h^2(t)\right),
\label{eq:alpha}
\end{equation}
    \item $\beta(t)$ is a time-dependent noise amplitude that measures the strength of the stochastic source term arising from the quantum nature of vacuum decay \cite{Starobinsky:1986fx, Martin:2011ib}. It depends on the background evolution as:
\begin{equation}
\beta(t) = \beta_0 \sqrt{\frac{16\pi\sigma}{3} h(t) \left(\frac{3}{8\pi\sigma} - h^2(t)\right)},
\end{equation}
where $\beta_0$ is a normalization constant, which will be adjusted to match the amplitude of perturbations.
    \item $\xi(t)$ is a Gaussian white noise term with zero mean, often normalized as $\langle \xi(t) \xi(t') \rangle = \delta(t - t')$, capturing the random energy-momentum transfer into the radiation bath \cite{Starobinsky:1986fx}.
\end{itemize}
\begin{figure}[h!]
    \centering
    \includegraphics[width=0.7\textwidth]{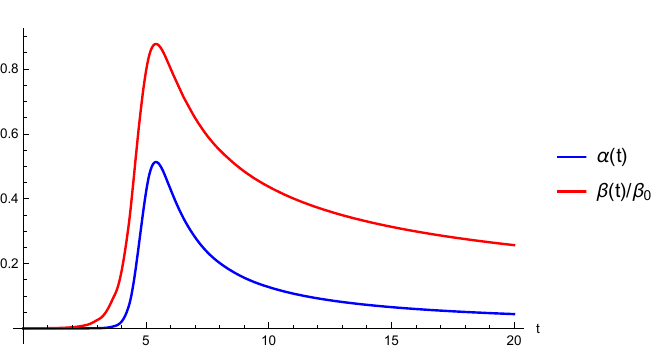}
    \caption{%
    \footnotesize
    Time evolution of $\alpha(t)$ (blue) and the normalized noise amplitude $\beta(t)/\beta_0$ (red) as functions of cosmic time $t$. Time is measured in units of the Planck time.%
    }
    \label{fig:alpha_beta}
\end{figure}

The time dependence of both $\alpha(t)$ and $\beta(t)$, shown in Figure~\ref{fig:alpha_beta}, reflects the changing cosmic medium, as both coefficients are governed by the same underlying vacuum decay dynamics. The damping encodes the suppression of perturbations via expansion, while the noise amplitude measures the fluctuation strength from the same decay process. This shared origin explains the proportionality between $\alpha(t)$ and $\beta^2(t)$.

\subsection{Physical Origin of Stochasticity and Coarse-Graining Prescription}
\label{sec:coarse_graining}

To clarify the open quantum system structure underlying Eq.~\eqref{eq:langevin}, we specify both the system-environment decomposition and the coarse-graining procedure explicitly.

\subsubsection{System and Environment}

The \emph{system} consists of the coarse-grained curvature perturbation $\mathcal{R}$ and the macroscopic radiation energy density---the cosmological observables evolving on scales $\gtrsim H^{-1}$. The \emph{environment} comprises microscopic quantum field degrees of freedom undergoing particle creation and annihilation during vacuum decay, specifically the short-wavelength modes of Standard Model fields at scales $\lesssim H^{-1}$ that mediate energy transfer from vacuum to radiation.

These environmental modes fluctuate rapidly on microscopic timescales but are not directly observable at cosmological scales. Tracing over (integrating out) these environmental degrees of freedom yields the effective stochastic description in Eq.~\eqref{eq:langevin}, with the noise term $\xi(t,\mathbf{x})$ encoding the residual quantum fluctuations from the traced-out environment. This procedure follows the open quantum systems methodology detailed in Ref.~\cite{Alicki:2023rfv} and reviewed in Sec.~\ref{sec:thermal_math}, where the Markovian master equation (Eq.~\ref{eq:MME}) governs the microscopic-to-macroscopic transition.

\subsubsection{Coarse-Graining Implementation}

The coarse-graining can be implemented through spatial or temporal averaging. We consider three possible prescriptions:

\textbf{Momentum Space Split (not used):} One could implement coarse-graining as a sharp cutoff in momentum space at the Hubble scale, with long-wavelength modes $k < aH$ constituting the system and short-wavelength modes $k > aH$ being integrated out. However, since all cosmologically relevant modes in our framework remain sub-Hubble ($k \gg aH$) throughout the transition (see Sec.~\ref{sec:horizon_dynamics}), such a split would place all observable scales in the ``environment'' rather than the ``system.'' We therefore adopt spatial/temporal averaging prescriptions that coarse-grain microscopic fluctuations within the sub-Hubble regime.

\textbf{Position Space Smoothing:} We average over spatial regions of characteristic size $\ell_{\rm CG} \sim H^{-1}$:
\begin{equation}
\bar{\mathcal{R}}(\mathbf{x},t) = \int d^3y\, W(|\mathbf{x}-\mathbf{y}|)\mathcal{R}(\mathbf{y},t),
\end{equation}
where $W(r)$ is a window function with width $\sim H^{-1}$ and $\bar{\mathcal{R}}$ denotes the coarse-grained curvature perturbation. This spatial averaging filters out fluctuations on sub-Hubble scales $\lesssim H^{-1}$, leaving only their cumulative macroscopic effect.

\textbf{Temporal Averaging (equivalent):} Since microscopic vacuum decay events occur on timescales $\tau_{\rm decay} \sim H^{-1}$ (as discussed in Sec.~\ref{sec:white_noise}), coarse-graining over timescales $\tau_{\rm CG} \sim H^{-1}$ averages out these rapid fluctuations, leaving only their cumulative effect as effective white noise. These two prescriptions are equivalent in the sense that they both implement coarse-graining at the Hubble scale, which is the natural separation between ``microscopic'' and ``macroscopic'' in cosmology.

\textbf{Key Distinction from Stochastic Inflation:} The crucial difference from stochastic inflation is that in our framework, modes remain sub-Hubble throughout ($k \gg aH$ for observable scales), so there is no super-Hubble freezing or horizon crossing that amplifies fluctuations. Instead, stochasticity arises from integrating out microscopic vacuum decay fluctuations on scales $\lesssim H^{-1}$ that continuously source energy-momentum transfer through the quantum-thermal processes described in Sec.~\ref{sec:thermal_math}. The coarse-graining timescale is $\tau_{\rm CG} \sim H^{-1}$, and fluctuations on shorter timescales appear as effective white noise to the coarse-grained dynamics, as justified in Sec.~\ref{sec:white_noise}.

The stochastic nature arises from quantum fluctuations associated with vacuum decay in the cosmic rest frame, where de Sitter space exhibits global thermal equilibrium at the Gibbons-Hawking temperature $T_{\rm dS} = h/(2\pi)$. This temperature characterizes the statistical properties of quantum field fluctuations for observers comoving with the FLRW expansion, providing a well-defined thermodynamic foundation for the decay processes. Unlike standard inflation, where quantum modes cross the horizon and freeze, here the stochasticity reflects the \textbf{statistical physics of continuous energy-momentum transfer} from the thermal vacuum state to radiation, mediated by the quantum decay processes.

Although horizon crossing plays a minimal role in this framework due to its brief duration, the cumulative effect of quantum fluctuations can be coarse-grained over physical volumes, leading to effective classical stochastic behavior \cite{Starobinsky:1986fx}. This justifies modeling the curvature perturbation using a classical stochastic differential equation, even though its source is fundamentally quantum. The process is inherently non-adiabatic due to entropy production, breaking the usual conservation of $\mathcal{R}$ on super-Hubble scales.

Curvature fluctuations are thus seeded not by a quantized inflaton field, but by the \textbf{statistical physics of quantum vacuum decay} in an evolving spacetime. This stochastic framework provides a physically transparent way to track the generation and amplification of primordial curvature perturbations. The key insight is that this \textbf{stochastic treatment} leads to perturbations directly tied to the quantum nature of the universe's early phases, allowing one to understand how features of the early Universe---such as the amplitude, tilt, and Gaussianity of fluctuations---may originate from this statistical process.

These characteristics provide an alternative to the standard inflationary paradigm, offering a potential new explanation for the observed features of the Cosmic Microwave Background (CMB) and large-scale structure, driven by the \textbf{quantum decay of vacuum energy} rather than inflaton field dynamics \cite{Martin:2011ib}.

\subsubsection{Non-Conservation of \texorpdfstring{$\mathcal{R}$}{R} and its Gauge-Invariant Status}

In conventional inflationary cosmology, the curvature perturbation $\mathcal{R}$ becomes conserved on super-Hubble scales in the absence of entropy production \cite{Mukhanov:1990me, Liddle:2000cg}. However, in the vacuum decay model considered here, horizon crossing is generically brief and plays a minimal role: while the comoving Hubble radius $(ah)^{-1}$ may decrease temporarily, this occurs only for short durations, and all physical scales remain sub-Hubble throughout the transition from de Sitter to radiation domination.

Consequently, the standard mechanism for the ``freezing'' of $\mathcal{R}$ does not apply. This reflects explicit time-translation symmetry breaking at the level of the effective dynamics, rather than spontaneous breaking associated with a single-clock background. Instead, perturbations evolve continuously under the influence of both deterministic background dynamics and stochastic fluctuations from vacuum decay. The propagation of modes is governed by the interplay between the expanding background and quantum fluctuations, with modes continuously evolving in a stochastic manner due to the non-adiabatic dynamics.

Despite this, $\mathcal{R}$ retains its usual gauge-invariant definition as the curvature perturbation on comoving hypersurfaces. Its evolution is governed by a stochastic differential equation, sourced by quantum fluctuations associated with vacuum decay. Since this decay injects both energy and entropy into the radiation bath, the system is inherently \emph{non-adiabatic}, and $\mathcal{R}$ evolves both deterministically and stochastically---not because of gauge artifacts, but as a direct result of real, physical energy-momentum exchange.

Therefore, the use of $\mathcal{R}$ as the primary scalar perturbation variable remains justified, even though its conservation is explicitly broken by the non-equilibrium dynamics of vacuum decay \cite{Liddle:2000cg}.

\subsubsection{Noise Structure: Additive vs.\ Multiplicative}

The additive noise structure in Eq.~\eqref{eq:langevin}, where $\xi$ enters independently of $\mathcal{R}$, reflects a specific assumption about the system-environment coupling. From an open quantum systems perspective, this corresponds to a bilinear interaction Hamiltonian of the form $H_I \sim \mathcal{R} \otimes O_E$, where $O_E$ represents environmental operators (e.g., energy-momentum fluctuations of short-wavelength modes). This coupling structure is appropriate when the stochastic source arises from {\em direct injection} of energy-momentum into the cosmological fluid, as in our vacuum decay scenario, rather than from noise in the equation of motion parameters themselves (which would yield multiplicative noise $\propto \mathcal{R}\cdot\xi$).

While multiplicative noise can arise in nonlinear theories through interactions beyond quadratic order between system and environment---indeed, this is generic once such interactions are present in any field theory, not specific to gravitational contexts---it typically emerges when environmental coupling modulates the effective parameters ($\alpha$, $\beta$) rather than directly sourcing $\mathcal{R}$. Nevertheless, we acknowledge this is a modeling assumption; a complete derivation from microscopic vacuum decay dynamics could in principle yield additional multiplicative corrections, which we defer to future work.

Regarding symmetry structure, in standard inflationary scenarios, time-translation invariance is spontaneously broken by the inflaton's classical trajectory, leading to an approximate shift symmetry $\mathcal{R} \to \mathcal{R} + \text{const}$. This implies interactions typically involve derivatives $\partial_t \mathcal{R}$ rather than $\mathcal{R}$ itself. In our framework, time-translation invariance is explicitly broken by the time-dependent background $h(t)$, and there is no conserved shift symmetry for $\mathcal{R}$---indeed, we explicitly state that $\mathcal{R}$ is {\em not} conserved due to non-adiabatic entropy production. The absence of a shift symmetry permits direct coupling to $\mathcal{R}$ in Eq.~\eqref{eq:langevin} rather than requiring derivative couplings. The stochastic source $\xi$ represents energy injection $\delta Q^0$, which couples to $\dot{\mathcal{R}}$ through the perturbed continuity equation (see Appendix B), yielding the form in Eq.~\eqref{eq:langevin} after integration.

\subsection{Justification of the White Noise Approximation}
\label{sec:white_noise}

The stochastic differential equation \eqref{eq:langevin} relies on modeling quantum fluctuations from vacuum decay as white noise $\xi(t)$ with $\langle \xi(t) \xi(t') \rangle = \delta(t - t')$. While this is standard in stochastic cosmology \cite{Starobinsky:1986fx}, its validity in our vacuum decay framework requires justification.

The white noise approximation is valid when the microphysical correlation time $\tau_{\rm decay}$ is much shorter than the coarse-graining time $\tau_{\rm CG}$ and the Hubble time $H^{-1}$: $\tau_{\rm decay} \ll \tau_{\rm CG} \ll H^{-1}$. Under this hierarchy, fluctuations decorrelate on the coarse-grained timescale, appearing as instantaneous random kicks.

In our model, vacuum energy decays through quantum processes characterized by the thermal scale $T_{\rm dS} = h/(2\pi)$, giving $\tau_{\rm decay} \sim (2\pi h)^{-1} \sim H^{-1}$. This suggests comparable timescales rather than clear separation, raising concerns about the approximation. However, the relevant comparison is with the damping timescale $\alpha(t)^{-1}$ of curvature perturbations. During the early de Sitter phase when $h(t) \approx h_0$, we find $\alpha(t)^{-1} \to \infty$ (extremely weak damping), while as the universe evolves toward radiation domination, $\alpha(t)$ increases and the damping timescale decreases to $\alpha^{-1} \sim H^{-1}$. For most of the evolution, $\alpha(t)^{-1} \gg \tau_{\rm decay}$, ensuring the noise appears effectively instantaneous compared to perturbation dynamics.

Following stochastic inflation \cite{Starobinsky:1986fx}, we implicitly coarse-grain over a smoothing window with width $\tau_{\rm CG}$ satisfying the hierarchy above. If microscopic noise has exponentially decaying correlations over $\tau_{\rm decay}$, the coarse-grained noise becomes delta-correlated provided $\tau_{\rm decay} \ll \tau_{\rm CG}$. In practice, solving the Langevin equation on timescales $\Delta t \sim H^{-1}$ implicitly performs this coarse-graining.

Our situation differs from stochastic inflation in two ways: we lack super-Hubble modes (all scales remain sub-Hubble, see Sec.~\ref{sec:horizon_dynamics}), and our noise source is quantum-thermal fluctuations in the vacuum decay rate itself rather than inflaton fluctuations. Nevertheless, the underlying logic is similar: if microscopic decay events decorrelate on timescales $\tau_{\rm decay} \lesssim H^{-1}$, and we coarse-grain on timescales $\sim H^{-1}$, the effective noise appears white. This is analogous to warm inflation \cite{Berera:1995wh}, where the condition $\Gamma_{\rm int} \gg H$ ensures fast thermalization, producing white noise. In our case, $\tau_{\rm decay} \sim H^{-1}$ places us on the boundary of validity.

Given the comparable timescales, corrections from \emph{colored noise} (finite correlation time $\tau_c$) may be important. Colored noise leads to non-Markovian dynamics with memory kernel $K(t, t')$ \cite{Vennin:2015,Pattison:2017}, potentially affecting the power spectrum amplitude, introducing scale-dependent corrections, generating running ($\alpha_s \neq 0$), or enhancing non-Gaussianity. Computing these corrections from the influence functional formalism \cite{Calzetta:2008} represents an important but technically challenging direction for future work.

We argue that white noise serves as a reasonable \emph{zeroth-order} description when: (i) $\tau_{\rm decay} \sim (2\pi h)^{-1}$ is at most comparable to $H^{-1}$, (ii) dynamics are coarse-grained over timescales $\sim H^{-1}$, and (iii) $\alpha(t)^{-1}$ varies from much larger to comparable to $H^{-1}$ during evolution. However, we emphasize important caveats: we have not performed explicit coarse-graining with a specified window function (our assumption is phenomenological), colored noise corrections are likely non-negligible given $\tau_{\rm decay} \sim H^{-1}$, and a complete justification requires specifying the microphysical decay mechanism and deriving the noise kernel from first principles.

The white noise approximation should be understood as a working hypothesis enabling tractable calculations while capturing essential physics. A complete treatment requires: (i) first-principles derivation of the noise correlation function from quantum field theory of vacuum decay in de Sitter space, (ii) quantifying colored noise effects on observables, and (iii) solving the full memory-kernel equation if significant. These represent important future directions that will either validate the approximation or reveal new phenomenology from memory effects. We have taken a first step by introducing spatial correlations in Sec.~\ref{sec:spatial_correlations}, demonstrating that realistic noise structures can generate observationally viable spectral tilts. Until such calculations are performed, our results should be understood as valid within the white-noise approximation, with potential $\mathcal{O}(1)$ corrections from colored-noise effects if $\tau_{\rm decay} \sim H^{-1}$.

\section{Horizon Dynamics and Scale Evolution}
\label{sec:horizon_dynamics}
\subsection{Absence of Horizon Crossing}
A crucial difference from standard inflation is that the comoving Hubble radius $(ah)^{-1}$ does not shrink significantly. The Hubble parameter $h(t)$ decreases monotonically, and unless initial conditions are fine-tuned very close to the critical value $h_c$, most modes either never exit the horizon or only briefly do so.

\begin{figure}[h!]
    \centering
    \includegraphics[width=0.7\textwidth]{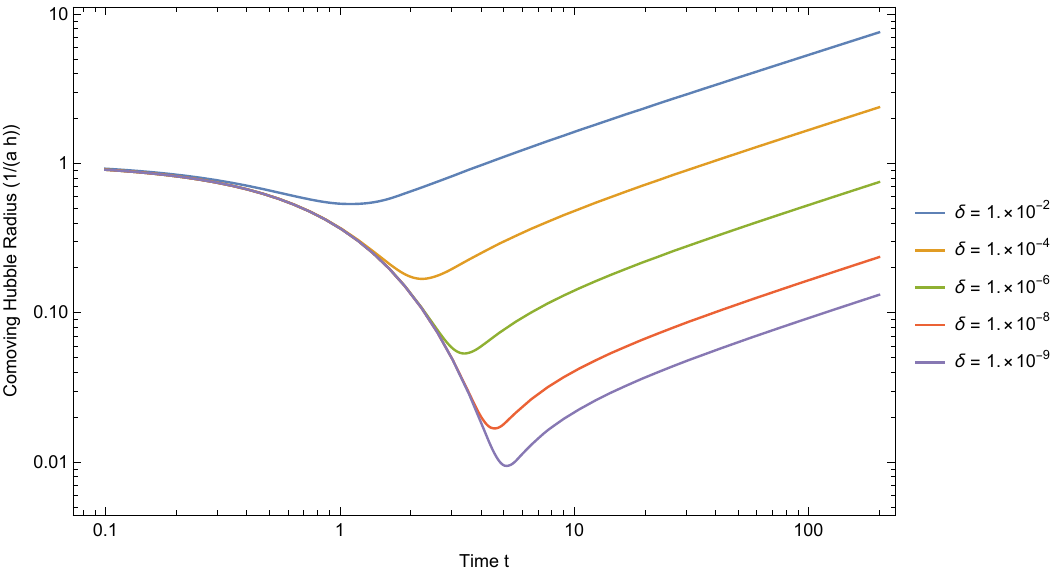}
    \caption{The comoving Hubble radius $1/(a h)$ is shown as a function of cosmic time $t$. The scale factor is normalized such that $a_0 = 1$ at $t=0$, and the initial Hubble parameter is chosen as $h_0 = \sqrt{\tfrac{3}{8 \pi \sigma}}\,(1 - \delta)$. The horizontal axis is given in units of the Planck time.}
    \label{fig:comoving}
\end{figure}

The duration of any transient shrinking phase is:
\begin{equation}
\Delta N(\delta) = \frac{1}{2}\ln\left(\frac{1-\delta}{1/\sqrt{2}}\right) - \frac{1}{4}\ln\left(\frac{\delta}{1-1/\sqrt{2}}\right) - \frac{1}{4}\ln\left(\frac{2-\delta}{1+1/\sqrt{2}}\right) 
\end{equation}
where $\delta = 1 - \frac{h_{\rm init}}{h_0}$ quantifies the fractional deviation of the initial Hubble parameter from the unstable equilibrium value $h_0 = \sqrt{\frac{3}{8\pi \sigma}}= \sqrt{2} \, h_c$. For small $\delta$,
$
\Delta N(\delta) \approx \frac{1}{128\delta}(1 - 4\delta + 2\delta^2),
$
but this overestimates the duration when $\delta \ll 10^{-3}$, since the exact behavior is only logarithmically divergent as $\delta \to 0$. In practice, the shrinking phase lasts less than one e-fold unless $\delta \lesssim 10^{-3}$, so achieving a prolonged shrinking phase requires significant fine-tuning. In most cases, the comoving Hubble radius either does not decrease much or only decreases for a very short period,  see Figure~\ref{fig:comoving} and Table~\ref{tab:efolds}.

\begin{table}[h!]
    \centering
    \begin{tabular}{|c|c|c|c|}
        \hline
        $\delta$ & $\Delta N$ (Exact) & $\Delta N$ (Approx) & Relative Error (\%) \\
        \hline
        $10^{-2}$      & 0.97 & 0.75 & 23 \\
        $10^{-4}$      & 2.13 & 78   & $3.6 \times 10^{3}$ \\
        $10^{-6}$      & 3.28 & $7.8 \times 10^{3}$ & $2.4 \times 10^{5}$ \\
        $10^{-8}$      & 4.43 & $7.8 \times 10^{5}$ & $1.8 \times 10^{7}$ \\
        $10^{-9}$      & 5.01 & $7.8 \times 10^{6}$ & $1.6 \times 10^{8}$ \\
        $10^{-15}$     & 8.46 & $7.8 \times 10^{12}$ & $9.2 \times 10^{13}$ \\
        $10^{-20}$     & 11.34 & $7.8 \times 10^{17}$ & $6.9 \times 10^{18}$ \\
        \hline
    \end{tabular}
    \caption{Comparison between the exact and approximate values of $\Delta N$ for different values of $\delta$. The last column shows the relative error in percent.}
    \label{tab:efolds}
\end{table}

\subsection{Extracting Observables: The Role of Evaluation Time}
Unlike inflation, where scale dependence arises from time-dependent horizon crossing of each mode, here the coefficients $\alpha(t)$ and $\beta(t)$ are \emph{independent of wavenumber} $k$. All modes evolve identically in time under the white noise assumption.

Therefore, the \emph{scale dependence} of the power spectrum does not originate from the evaluation time of the perturbations. Evaluating the power spectrum at any fixed cosmic time $t$ (e.g., the vacuum-radiation equality time $t_c$ defined by $\rho_{\rm dS}(t_c) = \rho_r(t_c)$) yields the \emph{same} scale dependence when spatial correlations are absent. This is because the coefficients governing the stochastic evolution are time-dependent but \emph{not scale-dependent}.

Physically, the time $t_c$ is chosen for convenience and conceptual clarity: it corresponds to the end of vacuum domination and the cessation of active stochastic driving. However, this choice does not affect the \emph{scale dependence} of the spectrum.

The stochastic differential equation for each Fourier mode is:
\begin{equation}
\dot{\mathcal{R}}_k(t) + \alpha(t) \mathcal{R}_k(t) = \beta(t) \xi_k(t)
\end{equation}
with $\xi_k(t)$ white noise uncorrelated in $k$.

The power spectrum at time $t_c$ is:
\begin{equation}
\mathcal{P}_{\mathcal{R}}(k, t_c) = \int_{t_i}^{t_c} dt' \beta^2(t') \exp\left[-2 \int_{t'}^{t_c} \alpha(s) ds\right]
\end{equation}

Since $\alpha(t)$ and $\beta(t)$ are independent of $k$, this implies $\mathcal{P}_{\mathcal{R}}(k) = \text{const.}$, i.e., a perfectly scale-invariant spectrum at any fixed time $t_c$ under the white noise assumption.

\section{Scale Dependence from Spatial Noise Correlations}
\label{sec:spatial_correlations}
\subsection{Beyond White Noise}
The above formulation assumes $\xi_k(t)$ is \emph{white noise}, meaning fluctuations are spatially uncorrelated at every point. This leads to a perfectly scale-invariant spectrum since the coefficients $\alpha(t)$ and $\beta(t)$ are independent of wavenumber $k$. However, this assumption is physically unrealistic.

Physically realistic sources of stochastic fluctuations, such as quantum vacuum effects or thermal radiation, inherently possess finite spatial correlations due to coherence over characteristic length scales. For example, thermal radiation fluctuations exhibit coherence lengths on the order of the thermal wavelength, while quantum vacuum fluctuations arising from vacuum decay processes display spatial correlations characterized by the underlying decay dynamics\cite{Boyanovsky:2006qi}.

We model these quantum or thermal fluctuations phenomenologically as classical stochastic noise with specified correlation properties, reflecting the underlying quantum-statistical physics in an effective classical framework \cite{Starobinsky:1986fx,Martin:2011ib, Calzetta:2008}. To incorporate these physically motivated correlations, we replace the white noise assumption with spatially correlated noise described by
\[
\langle \xi(\mathbf{x}, t) \xi(\mathbf{x}', t') \rangle = \delta(t - t') C(|\mathbf{x} - \mathbf{x}'|),
\]
where the spatial correlation function $C(r)$ encodes the coherence properties of the noise.

Under the white noise assumption, the perturbations are mode-independent. Introducing spatial correlations via a nontrivial $C(r)$ changes this: the noise power spectrum becomes $k$-dependent, imprinting scale-dependent features on the perturbations. This provides the key mechanism for generating observable spectral tilts.

\subsection{Power-Law Correlations}
\label{sec:correlations}
A simple yet instructive model for $C(r)$ is a power-law form
\[
C(r) =  r^\gamma,
\]
with a small positive exponent $\gamma$, indicating mild long-range correlations rather than perfect independence.\footnote{The power-law form $C(r) \propto r^\gamma$ is introduced here as a phenomenological ansatz. Its microphysical origin will be discussed in Sec.~\ref{sec:limitations}.}

To ensure causality is respected, the spatial correlation function $C(r)$ must be truncated or smoothly cut off beyond the horizon scale $r \gtrsim h^{-1}$, enforcing zero correlation at superhorizon distances. This cutoff does not significantly affect the phenomenology on observable subhorizon scales, but it guarantees that no unphysical correlations extend beyond the causal horizon, preserving the causal structure of the model.

Fourier transforming this spatial correlation function yields the noise power spectrum
\begin{equation}
P_\xi(k) = \int d^3r \, e^{-i \mathbf{k} \cdot \mathbf{r}} C(r) \propto k^{-(3+\gamma)},
\end{equation}
establishing a direct connection between the spatial correlation exponent $\gamma$ and the spectral tilt in momentum space.

\begin{figure}[h!]
    \centering
    \begin{subfigure}{0.48\textwidth}
        \centering
        \includegraphics[width=\textwidth]{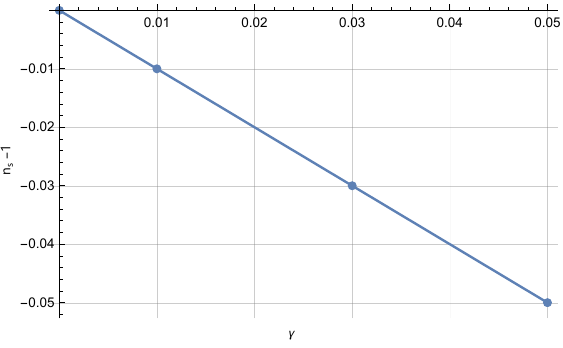}
        \caption{Spectral tilt $n_s - 1$ as a function of the spatial correlation exponent $\gamma$.}
        \label{fig:tilt_gamma}
    \end{subfigure}
    \hfill
    \begin{subfigure}{0.48\textwidth}
        \centering
        \includegraphics[width=\textwidth]{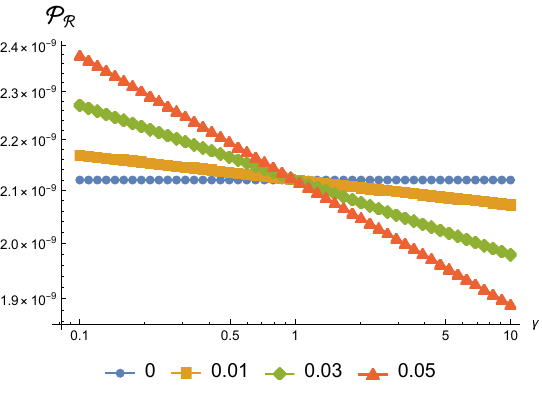}
        \caption{Power spectrum $\mathcal{P}_\mathcal{R}(k)$ for different values of $\gamma$. The horizontal axis shows $k$ in units of the pivot scale $k_*$.}
        \label{fig:PRk_gamma}
    \end{subfigure}
    \caption{Comparison of the spectral tilt (left) and the curvature perturbation power spectrum (right) for different spatial noise correlation exponents $\gamma$.}
    \label{fig:tilt_and_spectrum}
\end{figure}

This scale-dependent noise spectrum naturally imprints a scale-dependent amplitude on the curvature perturbations $\mathcal{R}_k(t)$, whose evolution is governed by
\begin{equation}
\dot{\mathcal{R}}_k(t) + \alpha(t) \mathcal{R}_k(t) = \beta(t) \sqrt{P_\xi(k)} \tilde{\xi}_k(t),
\end{equation}
where $\tilde{\xi}_k(t)$ is temporally white and spatially uncorrelated noise. Since the curvature power spectrum scales as
\begin{equation}
\mathcal{P}_{\mathcal{R}}(k) \propto k^3 P_\xi(k) \propto k^{3-(3+\gamma)} = k^{-\gamma},
\end{equation}
the spectral tilt is
\begin{equation}
n_s - 1 = \frac{d \ln \mathcal{P}_{\mathcal{R}}}{d \ln k} = -\gamma.
\end{equation}
Thus, by tuning the spatial correlation exponent $\gamma$, this framework can model a broad range of primordial spectral tilts, from blue-tilted spectra ($\gamma < 0$) to red-tilted ones ($\gamma > 0$). To match current observational constraints on the spectral index ($n_s \approx 0.965$), $\gamma$ should be approximately $0.035$, as illustrated in Figure~\ref{fig:tilt_gamma}.

Because the curvature power spectrum follows a pure power law in $k$, the spectral index $n_s$ is constant and does not run with scale. Therefore, the running of the spectral index
\begin{equation}
\alpha_s \equiv \frac{d n_s}{d \ln k} = 0
\end{equation}
in this simple power-law correlated noise model.

In summary, integrating physically motivated finite spatial coherence into the noise correlation function naturally connects the microphysics of quantum and thermal fluctuations to observable scale-dependent features in the primordial power spectrum.

In our phenomenological approach, the spectral tilt is determined by the spatial 
correlation exponent $\gamma$ in the noise kernel $C(r) \propto r^\gamma$, reflecting 
a fundamental difference from standard inflation where scale-dependence arises from 
time-dependent horizon crossing. Since the coefficients $\alpha(t)$ and $\beta(t)$ are 
independent of wavenumber $k$, scale-dependence must originate from spatial structure 
in the stochastic noise. This parallels how inflation requires specifying an inflaton 
potential $V(\phi)$ to predict $n_s$—both frameworks involve phenomenological inputs 
whose microscopic origin requires model-building at high energies. The physical origin 
of spatial correlations in our framework—whether from thermal coherence effects, finite 
decay rates, or influence functional memory kernels \cite{Calzetta:2008}—requires 
first-principles calculation from vacuum decay quantum field theory. Deriving $\gamma$ 
from microphysics represents an important direction for future work, analogous to 
deriving inflaton potentials from string theory or particle physics.
In this sense, near scale-invariance is not dynamically generated as in slow-roll inflation, but encoded phenomenologically in the long-range structure of the noise kernel.

\subsection{Physical Motivation for Correlations}
The spatial correlations can be understood through several physical mechanisms rooted in the quantum-thermal nature of vacuum decay in de Sitter space.

\subsubsection{Thermal Coherence and KMS Conditions}

In the vacuum decay scenario within de Sitter space, the radiation bath generated by the decay process is not simply a collection of uncorrelated particles. Instead, it emerges as a quantum-statistical ensemble with thermal characteristics determined by the Gibbons-Hawking temperature $T_{\rm dS} = h/(2\pi)$. This temperature sets the natural energy scale for the produced quanta, while the horizon scale $h^{-1}$ defines the characteristic spatial domain over which fluctuations may be correlated.

The quantum fields in de Sitter space satisfy the Kubo-Martin-Schwinger (KMS) condition in the cosmic rest frame, implying near-perfect thermal equilibrium at temperature $T_{\rm dS}$. Thermal equilibrium states typically exhibit correlations that decay exponentially beyond the thermal wavelength or correlation length, which here corresponds roughly to the Hubble horizon size $h^{-1}$.

\subsubsection{Finite Decay Rate Effects}
Unlike sudden reheating or instantaneous particle production, vacuum decay is a continuous, time-dependent process that injects energy and entropy into the radiation fluid \cite{Martin:2011ib, Calzetta:2008}. This continuous sourcing means that fluctuations in radiation energy density are not purely random but retain memory of previous decay events. The overlapping wave packets and coherent emissions associated with vacuum decay give rise to nontrivial temporal and spatial correlations, reflected in the small but positive scaling exponent $\gamma$.

However, the finite rate of vacuum decay prevents the noise from being purely white, thus generating scale-dependent correlations. Near the horizon scale, subleading quantum coherence effects and this finite decay rate induce deviations from pure white noise, producing correlations that grow slowly with distance on sub-horizon scales.

\subsubsection{Open Quantum Systems Perspective}
From the perspective of open quantum systems and the influence functional formalism in curved spacetime quantum field theory, the stochastic noise driving the radiation bath can acquire nonlocal structure both in space and time due to environmental memory effects \cite{Calzetta:2008,Martin:2011ib}. The noise kernel, which characterizes the correlation properties of the effective stochastic source, may therefore exhibit ``colored noise'' behavior rather than pure white noise.

A full first-principles calculation of this noise kernel in the vacuum decay scenario is challenging and beyond the present scope. However, calculations in related contexts—such as stochastic inflation \cite{Starobinsky:1986fx}, warm inflation with dissipative dynamics \cite{Berera:1995wh}, and influence functional treatments of particle creation in curved spacetime \cite{Calzetta:2008}—demonstrate that environmental integration can yield noise kernels with power-law or slowly decaying spatial correlations. While these examples do not directly apply to our scenario, they establish the conceptual plausibility that vacuum decay processes could generate non-trivial $C(r)$. We stress that deriving the specific form relevant to our model requires explicit calculation from the microphysics of de Sitter vacuum decay, which we leave to future work.

The correlation length is naturally limited by the horizon scale, ensuring that no unphysical correlations extend beyond the causal horizon while preserving the causal structure of the model.

\subsection{Horizon Problem and Statistical Homogeneity}
In this framework, the Universe begins in a high-entropy thermal de Sitter phase characterized by a Hubble parameter $h_0$, corresponding to a Gibbons-Hawking temperature $T_{\rm dS} = h_0/(2\pi)$. This temperature, as experienced in the cosmic rest frame, represents a global feature for comoving observers, ensuring that every comoving observer experiences the same thermal background.

Because of the strong symmetry properties of de Sitter space, the quantum vacuum state is
statistically identical across all causal patches \cite{BrosMoschella:1996, Marolf:2010nz} .
Every comoving region, though potentially causally disconnected in the classical sense, is drawn
from the same underlying quantum ensemble, ensuring that local observables such as the
energy--momentum tensor and correlation functions are homogeneous and isotropic throughout
space. When combined with the global satisfaction of the Kubo--Martin--Schwinger (KMS)
condition in the cosmic rest frame \cite{Alicki:2023rfv, Alicki:2023tfz},
this statistical uniformity acquires a genuine thermodynamic interpretation: for comoving
observers, the de Sitter vacuum behaves as a global thermal reservoir at the Gibbons--Hawking
temperature. As a result, the horizon problem is resolved not through causal equilibration driven
by accelerated expansion, but through the statistical equivalence imposed by the symmetries and
thermodynamic structure of the initial quantum state.

As the Universe evolves, vacuum energy decays into radiation in a process that occurs uniformly and locally, with each Hubble patch undergoing the same statistical evolution. This decay is accompanied by quantum fluctuations, which seed small inhomogeneities, but the overall process preserves large-scale isotropy because the underlying statistical ensemble is the same throughout space.

In this sense, causal equilibration through accelerated expansion is replaced by statistical equivalence arising from the symmetries of the initial quantum state.

\subsection{Addressing the Flatness Problem}
Closely related to homogeneity is the problem of spatial flatness, which this framework addresses in a similarly thermodynamic manner \cite{Marolf:2010nz}. The thermal de Sitter initial state naturally enforces spatial flatness due to its intrinsic symmetry. Unlike in standard inflation, where flatness is dynamically approached by rapid exponential expansion, here the Universe starts in a \emph{statistically homogeneous and isotropic} thermal equilibrium endowed with the de Sitter isometry group.

This symmetry dictates spatially flat slicings within each causal patch, fixing the spatial geometry at the level of the initial quantum-thermal ensemble. Consequently, spatial curvature is not a dynamical variable requiring dilution, but rather a fixed property of the initial state. Even accounting for the brief transient shrinking phase of the comoving Hubble radius, the overall spatial geometry remains effectively flat thanks to the global uniformity and symmetry of the thermal vacuum state.

Therefore, the flatness problem is resolved thermodynamically: flatness is not attained dynamically through inflationary expansion but emerges inherently from the initial quantum-thermal conditions of the Universe.

\subsection{Tensor Perturbations and the Tensor-to-Scalar Ratio}
\label{sec:tensors}

The tensor-to-scalar ratio $r$ is a key observable for distinguishing early universe models. In standard slow-roll inflation, tensor modes (primordial gravitational waves) are generated by quantum fluctuations of the metric itself during inflationary expansion, yielding $r = 16\epsilon$ where $\epsilon$ is the first slow-roll parameter. Here we examine whether tensor perturbations are generated in our vacuum decay framework.

Tensor perturbations $h_{ij}(t, \mathbf{x})$ represent transverse-traceless fluctuations in the spatial metric, evolving according to:
\begin{equation}
\ddot{h}_k + 3H\dot{h}_k + \frac{k^2}{a^2}h_k = S_k^{\rm tensor} = \frac{16\pi G}{a^2}\Pi_{ij}^{(TT)},
\label{eq:tensor_eom}
\end{equation}
where $\Pi_{ij}$ is the anisotropic stress tensor. Tensor modes are not generated spontaneously from a homogeneous, isotropic background---they require either quantum fluctuations of the metric itself during accelerated expansion, or anisotropic stress sources $\Pi_{ij} \neq 0$.

In standard inflation, quantum vacuum fluctuations in $\delta g_{\mu\nu}$ are promoted to classical perturbations through horizon crossing and squeezing during exponential expansion. Modes cross the horizon when $k = aH$, freeze on super-Hubble scales with amplitude $|h_k| \sim H_{\rm inf}/M_{\rm Pl}$, and re-enter during radiation/matter domination as classical gravitational waves. This mechanism relies fundamentally on prolonged horizon crossing.

Our vacuum decay model differs in three critical respects that suppress tensor generation at linear order:

\textbf{(1) No Prolonged Horizon Crossing:} As demonstrated in Sec.~\ref{sec:horizon_dynamics}, the comoving Hubble radius $(ah)^{-1}$ either never shrinks or shrinks only briefly ($\Delta N \lesssim 1$ e-fold for generic initial conditions, see Table~\ref{tab:efolds}). Without sustained horizon crossing, there is no mechanism to freeze quantum metric fluctuations and convert them to classical tensor perturbations.

\textbf{(2) Scalar Source, Not Metric Fluctuations:} Perturbations originate from energy-momentum fluctuations in the stochastic vacuum decay process, not from quantum fluctuations of the metric. The source term $\xi(t, \mathbf{x})$ in Eq.~\eqref{eq:langevin} represents fluctuations in energy injection rate $\delta Q^0 = \beta(t)\xi(t, \mathbf{x})$. This is a \emph{scalar} source coupling to curvature perturbations $R$ but not to transverse-traceless tensor modes $h_{ij}$. Energy density perturbations $\delta\rho$ source the $00$-Einstein equation $\nabla^2 R = 4\pi G a^2 \delta\rho$, while Eq.~\eqref{eq:tensor_eom} is sourced only by anisotropic stress.

\textbf{(3) Perfect Fluid with Zero Anisotropic Stress:} The radiation produced by vacuum decay is modeled as a perfect fluid with $w = 1/3$ and vanishing anisotropic stress $\Pi_{ij} = 0$, implying $S_k^{\rm tensor} = 0$. With no source term and modes remaining sub-Hubble, the tensor equation has no mechanism to generate perturbations. Any initial tensor perturbations decay as $h_k \sim a^{-1}$.

Combining these arguments, we conclude that \emph{at linear order in perturbation theory}, our vacuum decay model predicts:
\begin{equation}
r \approx 0.
\label{eq:r_zero}
\end{equation}
This is a sharp, testable prediction distinguishing our framework from single-field slow-roll inflation where $r = 16\epsilon > 0$ is generic. Current constraints place $r < 0.036$ (95\% CL) from Planck+BICEP/Keck \cite{BICEP:2021}, so our prediction is fully consistent with observations.

It is important to note that $r \approx 0$ holds only at \emph{linear order}. Nonlinear effects could potentially generate small tensor perturbations through scalar-scalar mode coupling at second order. The amplitude of such induced modes scales as $\mathcal{P}_h^{\rm induced} \sim \mathcal{P}_R^2 \times \mathcal{T}(k, \tau)$, analogous to secondary gravitational wave generation during radiation domination \cite{Mollerach:2003}. For $\mathcal{P}_R \sim 10^{-9}$, this gives $r^{\rm induced} \sim 10^{-9}$, far below current and future observational sensitivities. Additionally, if the radiation bath has non-zero viscosity or incomplete thermalization, small anisotropic stress $\Pi_{ij} \sim \eta \partial_i v_j$ could source tensors, but amplitudes depend on unspecified microphysical transport coefficients. We expect any nonlinear corrections satisfy $r^{\rm NL} \ll 10^{-3}$.

The prediction $r \approx 0$ provides a clear observational test: detection of $r > 10^{-3}$ would rule out our simplest linear model and favor inflationary scenarios, while non-detection down to $r \sim 10^{-4}$ would increasingly disfavor large-field inflation. Combined with the spectral tilt $n_s \approx 0.965$ accommodated through spatial noise correlations (Sec.~\ref{sec:spatial_correlations}), the absence of tensors would provide strong evidence for a non-inflationary origin of perturbations. Current and future CMB experiments (LiteBIRD, CMB-S4) aiming to probe $r \sim 10^{-3}$ will provide a crucial test of our framework.

\subsection{Comparison with Related Approaches}
\begin{itemize}
  \item \textbf{Warm Inflation (Berera et al.)} \cite{Berera:1995wh}:
  Warm inflation couples the inflaton field dynamically to a thermal bath maintained throughout inflation by dissipative processes. Fluctuations in warm inflation are driven by interactions between the inflaton and this thermal bath, leading to sustained nonzero temperature during inflation. In contrast, our model does not rely on a fundamental inflaton-dissipative coupling but instead considers vacuum decay itself as the energy source for the stochastic radiation bath, with correlations arising from the finite decay rate and continuous injection process.
\item \textbf{Stochastic Inflation (Starobinsky)} \cite{Starobinsky:1986fx}:
  Starobinsky's stochastic inflation formalism treats long-wavelength inflaton fluctuations as a classical stochastic process driven by quantum noise from short-wavelength modes crossing the horizon. This framework models the inflaton dynamics and metric fluctuations via a Langevin equation with white noise approximations. Our approach differs by focusing on the stochastic nature of vacuum decay and the induced radiation bath, rather than inflaton fluctuations, resulting in noise that is \emph{not} purely white but exhibits scale-dependent correlations due to the finite decay rate.
  \item \textbf{Decoherence of Cosmological Fluctuations (Calzetta \& Hu)} \cite{Hu:1992dc}:
  Calzetta and Hu analyze how interactions with environmental degrees of freedom lead to decoherence of primordial quantum fluctuations, facilitating their classicalization. Their open quantum systems approach elucidates environment-induced decoherence suppressing quantum interference in cosmological perturbations. While decoherence underpins classicality of fluctuations in both approaches, our model explicitly tracks how the vacuum decay process acts as a continuous source of noise and correlations, thereby providing a microscopically motivated stochastic environment rather than a generic decohering bath.
\end{itemize}

\section{Technical Considerations  }
\subsection{Backreaction Considerations}
In principle, backreaction effects may be a concern in this vacuum decay model because the stochastic curvature perturbations $\mathcal{R}(t)$ arise directly from quantum energy-momentum transfer during vacuum decay. Unlike in standard inflation, where super-Hubble modes decouple and evolve independently, here all physical scales remain sub-Hubble throughout the transition, and perturbations remain dynamically coupled to the background.

Furthermore, the system is inherently non-adiabatic due to continuous entropy production and energy injection, so nonlinear feedback from perturbations onto the background evolution could, in principle, accumulate over time. This raises the possibility that the effective equation of state or the Hubble parameter $h(t)$ might be modified by the perturbations themselves.

However, a naive estimate based on the variance $\langle\mathcal{R}^2\rangle$ shows that the amplitude of curvature perturbations remains small throughout the transition. Since $\mathcal{R}(t)$ evolves according to a linear stochastic differential equation with bounded noise amplitude, its fluctuations do not grow uncontrollably. As a result, the system stays well within the linear regime, and backreaction from perturbations is unlikely to significantly alter the background dynamics. This justifies treating the background evolution as governed by the deterministic equation for $h(t)$, without needing to account for higher-order feedback effects.

\subsection{Model Limitations and Open Questions}
\label{sec:limitations}

While our vacuum decay framework successfully addresses several cosmological puzzles and produces testable predictions, it is important to acknowledge that the present work establishes a \emph{phenomenological framework} rather than a complete microphysical theory. We outline the main limitations below.

\subsubsection{Frame-Dependent Thermal Interpretation}

Our framework relies on the thermal interpretation of de Sitter space from Ref.~\cite{Alicki:2023rfv}, which we have made mathematically explicit in Sec.~\ref{sec:thermal_math}. While the Markovian master equation (Eq.~\ref{eq:MME}), KMS condition (Eq.~\ref{eq:KMS}), and resulting thermal state (Eq.~\ref{eq:Gibbs}) were rigorously derived for constant-$h$ de Sitter space, our model involves time-dependent $h(t)$ during the vacuum decay transition. This raises several concerns:

(i) \textbf{Absence of global timelike Killing vectors:} In strictly de Sitter space with constant $h$, there exist global symmetries (the de Sitter isometry group) that underpin the thermal interpretation. When $h$ varies with time, these symmetries are broken, and the notion of a global thermal state becomes more subtle. The framework of Ref.~\cite{Alicki:2023rfv} was formulated for constant $h$, where a time-independent density matrix can be defined. With time-dependent $h(t)$, it is unclear whether a well-defined global Hamiltonian generating time translations exists in the usual sense.

(ii) \textbf{Applicability during transition phase:} The KMS condition (Eq.~\ref{eq:KMS}) and Stefan-Boltzmann scaling (Eq.~\ref{eq:StefanBoltzmann}) were derived for pure de Sitter space. During the transition when $h(t)$ varies, the spacetime is quasi-de Sitter, and the thermal properties may be modified. Our phenomenological application assumes these properties persist in some adiabatic or slowly-varying sense when $|\dot{h}/h^2| \ll 1$, but this is an extrapolation beyond the regime rigorously established in Ref.~\cite{Alicki:2023rfv}.

(iii) \textbf{Frame-dependence:} As emphasized in Sec.~\ref{sec:thermal_math}, the thermal interpretation is valid only in the cosmic rest frame. The KMS condition is not a covariant statement but a frame-dependent one, tied to the choice of time coordinate in the FLRW metric.

\subsubsection{Phenomenological Spatial Correlations}

The most significant limitation concerns the spatial correlation function $C(r) = r^\gamma$ introduced in Sec.~\ref{sec:spatial_correlations}. While this power-law form successfully generates observationally viable spectral tilts ($\gamma \approx 0.035$ for $n_s \approx 0.965$), it remains an \emph{ansatz} rather than a derived result. A complete theory requires deriving $C(r)$ from first principles using quantum field theory of vacuum decay in curved spacetime.

Several approaches could address this. The influence functional formalism \cite{Calzetta:2008} encodes noise correlations through the noise kernel obtained by integrating out short-wavelength modes. Alternatively, finite-temperature quantum field theory in de Sitter space, constrained by the KMS condition in the cosmic rest frame \cite{Alicki:2023rfv}, could calculate energy-momentum tensor correlation functions that directly source curvature perturbations. Until such calculations are performed, $C(r)$ should be regarded as a modeling choice parameterizing our ignorance of detailed microphysics, analogous to phenomenological inflaton potentials in standard inflation.

\subsubsection{Temporal Correlations and Microphysical Decay Mechanism}

As discussed in Sec.~\ref{sec:white_noise}, the white noise approximation is justified only when $\tau_{\rm decay} \ll H^{-1}$. Our estimate $\tau_{\rm decay} \sim H^{-1}$ suggests that temporal correlations (colored noise) may be important, leading to corrections to the power spectrum amplitude, potential running of the spectral index, and enhanced non-Gaussianity through memory effects. Computing these corrections requires deriving the memory kernel from the influence functional formalism, accounting for time evolution during vacuum decay.

More fundamentally, our model treats vacuum decay phenomenologically through the evolution equation \eqref{eq:hubble_evolution} without specifying the underlying mechanism. Key open questions include: which fields mediate the decay? What are the decay channels into Standard Model particles? What determines the decay rate parameter $\sigma$ from first principles? Why does decay proceed continuously rather than via bubble nucleation? 

Ideally, the mechanism should emerge from known physics. Possibilities include gravitational particle creation in time-dependent curved spacetimes \cite{Parker:1969,Birrell:1982}, the conformal anomaly driving energy transfer from vacuum to radiation \cite{Duff:1977}, or modifications to general relativity at high energies. Without specifying the microphysics, predictions for noise correlations, reheating temperature, and detailed thermal history remain incomplete.

\subsubsection{Energy Budget and Thermodynamic Consistency}

The energy budget in Eqs.~\eqref{eq:rho_dS} and \eqref{eq:rho_r} represents an effective description valid for cosmic rest frame observers experiencing global thermal equilibrium. The form $\rho_{\rm dS} \propto h^4$ reflects thermal scaling while maintaining the negative pressure ($w = -1$) characteristic of vacuum energy. The radiation component is not in equilibrium initially---equilibrium is approached asymptotically as vacuum decay completes. A fully consistent thermodynamic treatment requires specifying microphysical decay channels to calculate entropy production, deriving the effective equation of state from underlying field theory, and ensuring energy-momentum conservation including quantum corrections.

\subsubsection{Observational Predictions}

Our model makes robust predictions including zero tensor-to-scalar ratio ($r \approx 0$ at linear order, Sec.~\ref{sec:tensors}) and Gaussian perturbations ($f_{\rm NL} \approx 0$). 
However, the spectral tilt $n_s - 1 = -\gamma$ depends on the phenomenological parameter $\gamma$, analogous to how inflation's spectral tilt depends on the phenomenological inflaton potential $V(\phi)$. Deriving $\gamma$ from first principles represents an important goal for future work, as discussed in Sec.~5.2.
 The model should also produce isocurvature modes due to non-adiabatic evolution, which we have not calculated or compared with observational bounds. Additionally, we have not performed detailed likelihood analysis using CMB and large-scale structure data, leaving our claim of observational viability qualitative rather than quantitative.

At second order in perturbation theory, mode coupling between scalar perturbations could generate small contributions to non-Gaussianity and induced tensor modes, though we expect $r^{\rm NL} \ll 10^{-3}$ based on suppression by powers of $\mathcal{P}_R$.

\section{Conclusions}

We have presented a comprehensive analysis of cosmological perturbations in a vacuum decay model that offers an alternative to standard inflation. The framework generates primordial fluctuations through stochastic processes associated with quantum vacuum decay rather than inflaton dynamics.

Our analysis reveals a novel mechanism where perturbations arise from stochastic vacuum decay processes rather than the conventional horizon crossing dynamics of inflation. This approach naturally produces observable spectral tilts that emerge from spatial noise correlations, with the spectral index satisfying $n_s - 1 = -\gamma$, providing a direct connection between the microscopic decay physics and large-scale observational signatures.

The model successfully addresses fundamental cosmological problems through initial thermal equilibrium, resolving both the horizon and flatness problems without requiring fine-tuned initial conditions. A particularly distinctive feature is the prediction of zero tensor-to-scalar ratio, which provides a clear observational distinction from inflationary models and offers a definitive test of the framework. Despite the stochastic nature of the underlying process, the linear evolution of perturbations preserves Gaussianity regardless of spectral properties, ensuring consistency with current observational constraints.

The model demonstrates that viable alternatives to inflation can emerge from well-established physics---quantum field theory in curved spacetime and statistical mechanics---without requiring new fields or potentials. This represents a significant conceptual shift, suggesting that the complex inflaton dynamics traditionally invoked may not be necessary for generating the observed features of our universe.

While several aspects remain to be developed, particularly the first-principles derivation of spatial correlations, the framework provides a promising foundation for understanding early universe dynamics. Future investigations should focus on deriving noise correlations directly from fundamental vacuum decay processes, specifying the microphysical decay mechanism and its thermodynamic consistency, performing detailed observational comparisons including full likelihood analysis with CMB and large-scale structure data, exploring potential extensions that might generate small but non-zero tensor modes through second-order effects, and investigating possible non-Gaussian signatures arising from higher-order effects in the stochastic evolution.

This approach represents a significant departure from inflaton-based models while maintaining full compatibility with observations, suggesting that the physics of the early universe may be simpler and more fundamental than currently assumed. The framework opens new avenues for theoretical development and provides concrete predictions that can guide future observational programs in distinguishing between competing paradigms for the origin of cosmic structure.

\section*{Acknowledgements}
The author would like to thank Robert Alicki and Scott Dodelson  for insightful discussions and valuable comments. GB is supported by the Spanish grants  CIPROM/2021/054 (Generalitat Valenciana),  PID2023-151418NB-I00 funded by MCIU/AEI/10.13039/501100011033/, and by the European ITN project HIDDeN (H2020-MSCA-ITN-2019/860881-HIDDeN).

\appendix

\section{Gibbons-Hawking Temperature and the Cosmic Rest Frame}
The interpretation of de Sitter temperature requires careful consideration of observer frames. The Gibbons-Hawking temperature $T_{\rm dS} = h/(2\pi)$ appears universal, but its physical meaning depends crucially on the reference frame.

In the \textbf{cosmic rest frame}---defined by observers comoving with the FLRW expansion---the Kubo-Martin-Schwinger (KMS) condition is satisfied globally. This frame sees the universe as homogeneous and isotropic, with quantum fields in perfect thermal equilibrium at temperature $T_{\rm dS}$.

In contrast, \textbf{static patch} observers experience only local thermal equilibrium with position-dependent temperature due to gravitational redshift. Their vacuum state is thermal in an observer-dependent sense but lacks global equilibrium.

Only cosmic rest frame observers experience a genuinely thermal universe in the full statistical mechanical sense, justifying our treatment of de Sitter space as an initial thermal state for cosmological evolution.

\section{Derivation of the Stochastic Evolution Equation}

We derive the stochastic differential equation governing the curvature perturbation $\mathcal{R}(t,\mathbf{x})$ in comoving gauge. Throughout the vacuum decay transition, all cosmologically relevant modes satisfy $k \gg ah$, remaining well within the Hubble radius.

We begin with energy-momentum conservation in the presence of a stochastic source:
\begin{equation}
\nabla_\mu T^{\mu\nu}_{\rm fluid} = Q^\nu,
\end{equation}
where $Q^\nu$ describes the energy-momentum injected into the fluid by vacuum decay. The background spacetime is a spatially flat FLRW universe with metric:
\begin{equation}
ds^2 = -dt^2 + a(t)^2 d\mathbf{x}^2.
\end{equation}

At the background level, the energy density $\rho(t)$ and pressure $p(t)$ of the fluid obey:
\begin{equation}
\dot{\rho} + 3h(\rho + p) = Q^0,
\end{equation}
where \( h = \dot{a}/a \) is the Hubble parameter and \( Q^0 \) is the background energy injection rate.

We work in comoving gauge, where the fluid velocity perturbation vanishes: \( \delta u = 0 \). The perturbed metric takes the form:
\begin{equation}
ds^2 = -(1+2\Phi) dt^2 + a(t)^2 (1+2 \mathcal{R}) \delta_{ij} dx^i dx^j.
\end{equation}
Here, the curvature perturbation \(\mathcal{R}(t,\mathbf{x})\) directly characterizes fluctuations on comoving hypersurfaces.

Stochastic fluctuations in the energy injection rate are modeled by:
\begin{equation}
\delta Q^0(t,\mathbf{x}) = \xi(t,\mathbf{x}),
\end{equation}
where \(\xi\) is a stochastic source representing microscopic vacuum decay events.

The perturbed continuity equation, including this stochastic source, reads:
\begin{equation}
\delta \dot{\rho} + 3h (\delta \rho + \delta p) - 3(\rho + p) \dot{\mathcal{R}} = \xi(t,\mathbf{x}).
\end{equation}

To proceed, we decompose the pressure perturbation into adiabatic and non-adiabatic parts:
\begin{equation}
\delta p = c_s^2 \delta \rho + \delta p_{\rm nad},
\end{equation}
where \( c_s^2 = \dot{p}/\dot{\rho} \) is the adiabatic sound speed, and \(\delta p_{\rm nad}\) captures entropy (non-adiabatic) fluctuations, including those induced by vacuum decay.

Substituting this into the perturbed continuity equation yields:
\begin{equation}
\delta \dot{\rho} + 3h(1 + c_s^2)\delta \rho - 3(\rho + p)\dot{\mathcal{R}} = \xi + 3h \delta p_{\rm nad}.
\end{equation}

On sub-Hubble scales (\( k \gg a h \)), pressure gradients drive rapid oscillations in \(\delta \rho\) and \(\mathcal{R}\). Using Einstein's equations (not shown here in detail), one finds that the curvature perturbation obeys a second-order evolution equation of the form:
\begin{equation}
\ddot{\mathcal{R}} + \left(3h + \frac{\dot{p}}{\rho + p} \right) \dot{\mathcal{R}} + \frac{c_s^2 k^2}{a^2} \mathcal{R} = \frac{h}{\rho + p} \left( \delta p_{\rm nad} + \xi \right).
\end{equation}

The damping term above is exact. However, if the equation of state \( w(t) = p(t)/\rho(t) \) varies slowly on sub-Hubble timescales (i.e., \( \dot{w}/w \ll c_s k/a \)), we can approximate it as constant over an oscillation period. In that case:
\[
3h + \frac{\dot{p}}{\rho + p} \approx 3h(1 + w),
\]
restoring the standard friction term for acoustic oscillations in a fluid with constant equation of state.

In this regime, the gradient term \( (k^2/a^2)\mathcal{R} \) dominates the dynamics, leading to high-frequency oscillations in \(\mathcal{R}\). To isolate the slow evolution of the oscillation envelope, we coarse-grain over timescales long compared to \( \omega_k^{-1} = a/(c_s k) \). This procedure yields an effective first-order Langevin-type equation for the coarse-grained curvature perturbation:
\begin{equation}
\dot{\mathcal{R}} + \alpha(t)\mathcal{R} = \beta(t)\xi(t,\mathbf{x}),
\end{equation}
where the damping coefficient is \( \alpha(t) = 3h(1 + w) \), and the coupling \( \beta(t) \) reflects the influence of stochastic energy injection. Since the source \( \xi \) arises from fluctuations in the vacuum decay rate, which is tied to the background evolution, it is natural to model \( \beta(t) \) as a function of the evolving background.

\bibliographystyle{apsrev4-2}  

\bibliography{vacuum}     

\end{document}